\documentclass[article,accept,moreauthors,preprints]{Definitions/mdpi} 

\usepackage[utf8]{inputenc}
\usepackage[english,greek,main=english]{babel}
\usepackage{alphabeta}

\newcommand{\secref}[1]{\S~\ref{#1}} 
\newcommand{\myitem}[1]{\vspace{0.25\baselineskip}\noindent\textbf{#1}}
\newcommand{\rev}[1]{\textcolor[rgb]{0.00,0.00,0.00}{#1}}
\newcommand{\revv}[1]{\textcolor[rgb]{0.00,0.00,0.00}{#1}}

\firstpage{1} 
\pubvolume{1}
\issuenum{1}
\articlenumber{0}
\pubyear{2023}
\copyrightyear{2023}
\datereceived{ } 
\daterevised{ } 
\dateaccepted{ } 
\datepublished{ } 
\hreflink{https://doi.org/} 

\Title{Analyzing large scale political discussions on Twitter: the use case of the Greek wiretapping scandal (\#ypoklopes)}

\TitleCitation{Analyzing large scale political discussions on Twitter: the use case of the Greek wiretapping scandal (\#ypoklopes)}

\Author{Ilias Dimitriadis $^{\dagger}$, Dimitrios P. Giakatos $^{\dagger}$, Stelios Karamanidis $^{\dagger}$, Pavlos Sermpezis $^{\dagger}$,\\Kelly Kiki $^{\ddagger}$ and Athena Vakali $^{\dagger}$}

\AuthorNames{Ilias Dimitriadis, Dimitrios P. Giakatos, Stelios Karamanidis, Pavlos Sermpezis, Kelly Kiki and Athena Vakali}

\AuthorCitation{Dimitriadis, I.; Giakatos, D.P.; Karamanidis, S.; Sermpezis, P.; Kiki, K.; Vakali, A.}

\address{%
$^{\dagger}$ Aristotle University of Thessaloniki, Greece; \{idimitriad, dgiakatos, skaraman, sermpezis, avakali\}@csd.auth.gr\\
$^{\ddagger}$ iMEdD (incubator for Media Education and Development), Greece; k.kiki@imedd.org}

\abstract{In this paper, we study the Greek wiretappings scandal, which has been revealed in 2022 and attracted a lot of attention by press and citizens. Specifically, we propose a methodology for collecting data and analyzing patterns of online public discussions on Twitter. We apply our methodology to the Greek wiretappings use case, and present findings related to the evolution of the discussion over time, its polarization, and the role of the media. The methodology can be of wider use and replicated to other topics. Finally, we provide publicly an open dataset, and online resources with the results. }

\keyword{data journalism; data analysis; visualization} 

\begin{document}

\section{Introduction}\label{sec:intro}

A prolonged monitoring of the mobile phones of journalists and politicians has been revealed in 2022 and shook the Greek political scene to its core. The scandal, commonly known as the Greek Wiretapping Scandal, has been in the public sphere for more than a year, it attracted a lot of attention from Greek and international media~\cite{schmitz2022wiretapping, aljazeera2022why, reuters2023why, smith2023greek, partsakoulaki2023how}, and raised a lot of discussions on social media.

In particular, the hashtags \#υποκλοπες and \#ypoklopes (the words in Greek and "Greeklish" for wiretapping) were among the top so-called "trending topics" on Greek language Twitter for several months. 
The importance of the topic and the large public interest motivated a data journalism effort for tracking, mapping, characterizing, and analyzing the public discussions on Twitter about the scandal. Tracking online discussions has been proven useful in a number of cases for extending insights on public issues~\cite{alshehhi2019cross, alsinet2017weighted, bruns2012researching}

\begin{figure}
  \includegraphics[width=\linewidth]{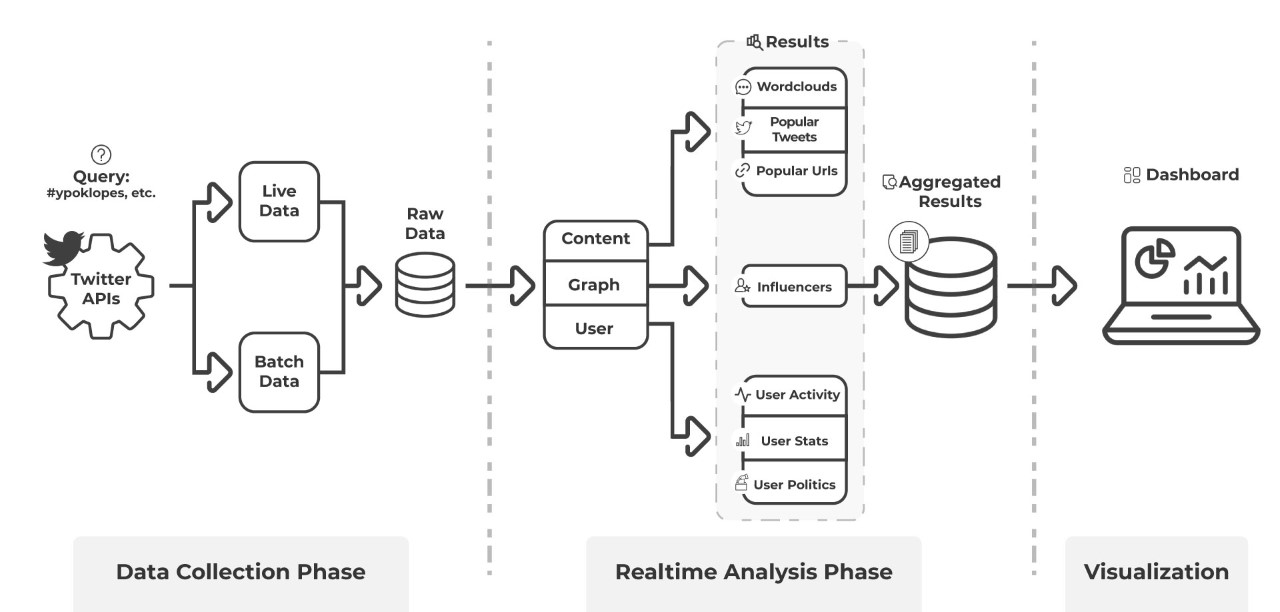}
  \caption{Overview of the methodology.}
  \label{fig:working_methodology}
\end{figure}

\rev{The contributions of this paper are threefold:} we present the methodology, dataset and results of this effort. Despite the fact that our main focus is on the Greek wiretapping scandal, our methodology is applicable to other use cases of online discussions as well, and reveal insights and trends. Specifically, our contributions can be summarized as follows:

\begin{itemize}[leftmargin=*,nosep]
    \item We design a methodology for monitoring and analyzing large scale political discussion on Twitter (\secref{sec:data} and \secref{sec:methodology}). The methodology is generic (i.e., can be replicated to other use cases) and includes the steps of data collection, political inference, bot detection, polarization quantification, and analysis of users and content. An overview of the methodology is depicted in Fig.~\ref{fig:working_methodology}.
    \item We study the Greek wiretappings scandal (see details in \secref{sec:scandal-history}), and collect a dataset of the \textit{entire} discussion on Twitter with a duration longer than a year (\secref{sec:dataset}). Moreover, we compile a number of complementary datasets related to political attributions, accounts of media, and Twitter bot accounts (\secref{sec:complementary-datasets}). We publicly share these datasets in~\cite{dimitriadis2023wiretappingdataset}.
    \item We present and discuss the findings of our analysis (\secref{sec:analysis}). For example, we show how the volume of tweets changes over time and significantly intensifies upon major events or news publications (\secref{sec:quant-analysis}), we analyze the role of media as main drivers and influencers of the online discussion (\secref{sec:media-role}), we quantify the participation of users attributed to the political "Left" and "Right" in the discussion (\secref{sec:left-right}) and how polarized they are (\secref{sec:polarization-analysis}).
\end{itemize}
Supplementary to this paper, all the results of our analysis can be accessed in an online Web portal~\cite{portal2023ypoklopes}, and an extended critical discussion of the results can be found in the online article~\cite{kiki2023ypoklopes}.

\section{The Greek wiretapping scandal}\label{sec:scandal-history}

The Greek wiretapping scandal has been in the public sphere for 16 months. 
For a better understanding of the scope and results of our study, we provide the background and history of the scandal. In the following list we enumerate in chronological order a list of main events and news publications related to the wiretapping scandal:

\begin{itemize}[leftmargin=*,nosep]
    \item \textbf{Nov. 14, 2021}: the newspaper "Efimerida ton Syntakton" reports that the National Intelligence Service (NIS) is monitoring citizens, among who a Greek journalist covering the story of a 12-year-old refugee from Syria    \footnote{\url{https://www.efsyn.gr/themata/thema-tis-efsyn/319063_polites-se-kathestos-parakoloythisis-apo-tin-eyp}}
    . 
    \item \textbf{Nov. 16, 2021}: Two days later, the reporter and member of the journalist network "Reporters United", Stavros Malihoudis, publishes an article entitled "I am the journalist under surveillance by the NIS"\footnote{\url{https://www.reportersunited.gr/6976/eimai-o-dimosiografos-poy-parakoloythei-i-eyp/}}, explaining that he became aware of the fact by the aforementioned story
    \item \textbf{Dec. 16, 2021}: Two studies are published by the University of Toronto's Citizen Lab\footnote{\url{https://citizenlab.ca/2021/12/pegasus-vs-predator-dissidents-doubly-infected-iphone-reveals-cytrox-mercenary-spyware/}} and Meta\footnote{\url{https://about.fb.com/news/2021/12/taking-action-against-surveillance-for-hire/}}, on the unknown at the time Predator spyware, with its clients possibly extending to Greece. 
    \item \textbf{Jan. 2022}: The journalism groups "Reporters United"\footnote{\url{https://www.reportersunited.gr/7359/parakoloythiseis-eyp-siopi-o-vasilias-akoyei/}} 
    and "Inside Story"\footnote{\url{https://insidestory.gr/article/neo-logismiko-kataskopeias-predator-kai-oi-doyleies-stin-ellada}} 
    published investigations into the passing of an amendment in the Greek parliament on March 31, 2021 that altered the rules for the lifting of confidentiality of communications in Greece and for Predator and "business in Greece"
    , respectively. In the months that followed, the two groups were at the forefront of the journalistic investigation into the issue of surveillance and wiretapping, which however took a long time to start being extensively covered by the mainstream media and -consequently?- widely discussed by citizens.
    \item \textbf{Apr. 6, 2022}: The journalist Thanasis Koukakis files a complaint with the Hellenic Authority for Communication Security and Privacy (ADAE), requesting an investigation into the case of infection of his cell phone with Predator spyware. In the following days, three related journalistic investigations are published. 
    \item \textbf{Apr. 11-15, 2022}: Journalists Tasos Telloglou and Eliza Triantafyllou publish two related investigations in Inside Story, under the headlines “Who was monitoring journalist Thanasis Koukakis’s cell phone?” and “Koukakis surveillance case: The state knows” on April 11 and 14 respectively. Also, on April 15, journalists Nikolas Leontopoulos and Thodoris Chondrogiannos published an investigation in Reporters United, according to which the government was monitoring journalist Thanasis Koukakis. In the meantime there has been a statement by the Deputy Minister to the Prime Minister and Government Spokesperson, Yiannis Economou, where he had referred to the Koukakis case as surveillance by a private individual, stating among others that “obviously it is unthinkable in a country like Greece, under the rule of law, for any private individual to be able to monitor another private individual” –a statement that Thanasis Koukakis himself had commented on Twitter.
    \item \textbf{May 19, 2022}: Google’s Threat Analysis Group announces their assessment, put forth "with high confidence", that government-backed actors in at least eight countries, including Greece, have obtained exploit software.
    \item \textbf{July 26, 2022}: Nikos Androulakis, president of the socialists' political party (PASOK) and MEP, reports the attempted wiretapping of his cell phone.
    \item \textbf{Aug. 4, 2022}: Publication of a journalistic investigation by "Reporters United" and "Efimerida ton Syntakton", according to which transaction of Grigoris Dimitriadis (Secretary General of the Prime Minister) are linked to a former manager of the company Intellexa which markets the Predator spyware.
    \item \textbf{Aug. 5, 2022}: Resignation of Grigoris Dimitriadis and the head of the National Intelligence Service (NIS) Panagiotis Kontoleon.
    \item \textbf{Aug. 26, 2022}: Debate in Parliament on the wiretapping case.
    \item \textbf{Sept. 10, 2022}: A Greek parliament member (with the party of SYRIZA) reports the attempted interception of his mobile phone.
    \item \textbf{Nov. 7, 2022}: Publication of the first list of public figures allegedly under surveillance by the newspaper "Documento".
    \item \textbf{Dec. 8, 2022}: Debate on the new draft law on the NIS at the Plenary Session of the Parliament.
    \item \textbf{Jan. 10, 2023}: The Prosecutor of the Supreme Court, opines that Hellenic Authority for Communication Security and Privacy is not responsible for handling citizens' requests for information on whether they have been placed under surveillance and cannot address mobile telephony providers in this regard.
\end{itemize}

\begin{figure*}
  \includegraphics[width=\textwidth]{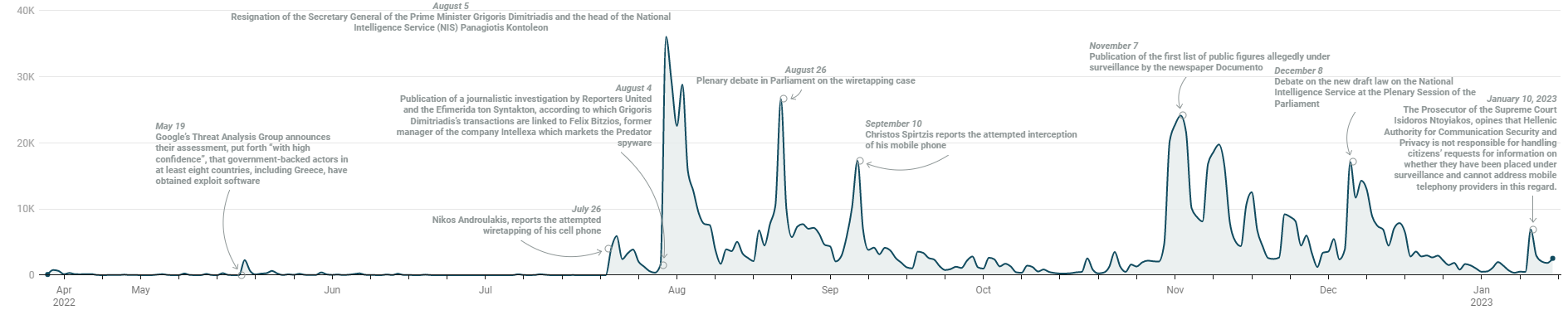}
  \caption{Total posts (including tweets, retweets, quotes, and replies) over the period of study (April 1, 2022 - January 14, 2022). A selection of important events and publications for the wiretapping scandal are annotated.}
  \label{fig:total_posts}
\end{figure*}

The scandal took a long time to start being extensively covered by the mainstream media and -consequently- widely discussed by citizens. As it can be seen in Fig.~\ref{fig:total_posts}, where we annotate the above events along with the total number of relevant tweets, the main discussion in Twitter started only on Aug. 2022.


\section{Data}\label{sec:data}

\subsection{Twitter API}
Twitter provides access to public data in the platform (user profiles, tweets, etc.) through an official API (v2)~\cite{twitter2023api} that can be used by all registered users. The API includes the Search API, which can be used to retrieve historical data, and the Streaming API, which can collect real-time data.\footnote{\rev{Since the time the survey was conducted (2022-2023), there have been significant changes in the costs incurred with accessing the Twitter API; this would differentiate the costs of the data collection phase in our methodology today.}}

To narrow down the collection of tweets on a specific topic, a set of keywords and/or hashtags can be given as input in the API call. The returned tweets will need to contain at least one of the terms of the given set. In our study, we compile the \#ypoklopes dataset by using the keywords and hashtags as described in~\secref{sec:dataset}.

\textit{Remark:} The Twitter API has some restrictions, e.g.,in terms of volume of data that can be collected. In our case, we used an academic licence that enabled us to retrospectively retrieve the entire volume of tweets relevant to the targeted discussion.

\subsection{The \#ypoklopes dataset}\label{sec:dataset}
\myitem{Keywords and Hashtags.} To collect the desired data, we selected a set of keywords and hashtags of \textit{terms} (e.g., "wiretapping" or "spyware") and \textit{names} (of journalists or politicians) related to the scandal. The detailed list is given in Table~\ref{tab:keywords}.

The selection of these keywords was based on journalistic and technological criteria; after a thorough initial listing of terms relevant to the conversation about the wiretapping case on Twitter, the available data were quantified and subjected to quality control. For example, while the acronym "ΕΥΠ" (National Intelligence Service or "NIS") was initially included in the terms, we found that our searches with the term "ΕΥΠ" would mainly return data that was not relevant to the subject. Therefore, it was eventually not included in the data collection terms.

\rev{Based on specific events during the tapping scandal period the initial set of hashtags was enriched with new ones, solely for the period of each event, but the main set of tracked hashtags remained the same. The web application~\cite{dimitriadis2023wiretappingdataset} (\url{https://ypoklopes.csd.auth.gr}) that provides complementary support to this research, allows for the identification of such events and their corresponding hashtags.}

Similarly, also for reasons related to ensuring the relevance of the data to the subject
: (i) the hashtag \#ανδρουλακης, which refers to the member of the EU parliament (MEP) and President of PASOK-KINAL, Nikos Androulakis, was added to the data collection criteria for tweets posted from July 20, 2022 (previous mentions of him on Twitter relate mainly to his activity as a MEP); (ii) all hashtags corresponding to the names of individuals are included in the data collection criteria for tweets posted until November 28, 2022.


\begin{table}
    \caption{Keywords and hashtags (denoted with \#) used as filter in the Twitter API requests.}
    \label{tab:keywords}
    \centering
    \begin{small}
    \begin{tabular}{l|l}
        \textbf{Keyword/hashtag} & \textbf{Comments} \\
        \hline
        \hline
         υποκλοπές & Words (in Greek or "Greeklish") \\
         \#υποκλοπες &  for "wiretapping", "surveillance", etc.\\
         \#υποκλοπές &\\
         υποκλοπη &\\
         \#παρακολουθήσεις &\\
         \#ypoklopes &\\
         \hline
         \#watergate & Other commonly used terms\\
         greekwatergate & for the topic\\
         \hline
         predator & Terms related to the software\\
         \#predator & used for the wiretappings\\
         \#predatorgate &\\
         \#pega &\\
         \#spyware &\\
         \hline
         \#δημητριαδης & Political figures\\
         \#κοντολεων & involved in the scandal\\
         \hline
         \#κουκακη & Journalist (under surveillance)\\
         \hline
         \#ανδρουλακης & Politician (under surveillance)        
    \end{tabular}
    \end{small}
\end{table}

Finally, we keep only tweets in Greek.

\myitem{Collection period.} We collected data since the beginning of 2022. In the first quarter of 2022 there was no conversation on Twitter, but for a few -not very relevant- tweets. Hence, we restrict our to the main corpus of tweets, starting on April 1, 2022. This time point corresponds to the publication of Thanasis Koukakis' monitoring case and related journalistic revelations.

\subsection{Complementary datasets}\label{sec:complementary-datasets}

\myitem{Political parties and politicians.} For analyzing political aspects of the discussion (see \secref{sec:analysis}), we collected all the accounts of Greek parties, members of the Greek parliament (MPs) and Greek members of the European parliament (MEPs) on Twitter using again the Twitter API. We retrieve the official Twitter usernames of Greek MPs, MEPs, and Parties using the relevant data on Vouliwatch~\cite{vouliwatch2023portal}, an independent, non-profit open governance initiative, and the European Parliament's website~\cite{europa2023europarlmeps}. 

The six parties participating in the Greek Parliament (2019-2023) are: New Democracy, SYRIZA, PASOK-KINAL, KKE, MeRA25, and Greek Solution. The government party (New Democracy) and "Greek Solution" are generally considered to be positioned in the political "Right", while the parties SYRIZA, KKE, and MeRA25 are in the political "Left", and PASOK-KINAL in the political "Center". We annotate the accounts of the parties and the affiliated politicians in the categories "Right", "Left", "Center", respectively.

\myitem{Media and journalists.} We wanted to characterize the activity and role of the media and journalists in the public discussion in Twitter. Due to the lack of relevant data sources, we had to rely on manual annotation of these accounts. However, given that in the collected dataset we identified more than 33K unique users, a manual annotation for all of them is infeasible.
Hence, we consider a set of the most "prevalent" unique accounts, i.e., the top 500 accounts in each of the following categories: those who (i) posted the most tweets, (ii) responded the most to third-party tweets, (iii) posted the most quotes, (iv) were quoted the most by others, and (v) were the most influential in the discussion about wiretapping (further discussed in~\secref{sec:influencers}). We end up with a set of 2262 unique users, out of which 407 users (18\%) have been identified as "Media/Journalists" (including mass media, newsrooms, journalists, or blogs). 

\myitem{Twitter bots.} Taking into account the evident presence of bots (i.e., automated accounts)~\cite{cresci2020decade} in Twitter and their impact, especially on political discussions~\cite{ferrara2020bots,ratkiewicz2011detecting}, we investigated which of the most prevalent accounts can be classified as bots. To this end, we used an online state-of-the-art tool for automated bot detection, namely Bot-Detective~\cite{kouvela2020bot, dimitriadis2021social}. 
In contrast to other political discussions and analyses on Twitter~\cite{portal2023ypoklopes}, only a minor percentage (1.4\%) of accounts were identified as actual bots.  

\myitem{Organizations and individuals.} Again, manually, we annotated as "Organizations" the accounts corresponding to organizations (excluding media and political parties), brands, etc. Finally, the remaining accounts were annotated as "Invididuals".

The detailed statistics of the account types (among the most prevalent accounts) in our datasets are given in Table~\ref{user_table}.

\begin{table}
\centering
\caption {Distribution of the most "prevalent" users.}
\label{user_table}
\begin{small}
\begin{tabular}{l |r |r}
 \textbf{Category} & \textbf{Number} & \textbf{Percentage} \\ [0.5ex] 
 \hline\hline
Individuals & 1688 & 74.6\%  \\ 
 \hline
Media/Journalists & 407 & 18.0\%  \\
 \hline
Political Accounts & 97 & 4.3\%  \\
 \hline
 Organizations & 38 & 1.7\% \\
 \hline
 Bots & 32 & 1.4\% 
\end{tabular}
\end{small}
\end{table}


\section{Methodology: Users and Network Characteristics}\label{sec:methodology}

Apart from quantitative metrics (e.g., volume of tweets per day, or most active users) that can be calculated directly from the collected dataset (\secref{sec:dataset}), for a deeper understanding of the phenomena taking place on the online discussion, we need to analyze also the user interactions and their opinions (microscopic level), as well as the macroscopic characteristic of the entire social network of users participating in the discussion.

In this section, we present our methodology for mapping user interactions to graphs (\secref{sec:graph-generation}), inferring the political attribution of users (\secref{sec:political-inference}), quantifying the polarization of the network (\secref{sec:polarization}), and identifying the most influencing users (\secref{sec:influencers}).

\subsection{Graph generation}\label{sec:graph-generation}

\begin{figure}
  \includegraphics[width=0.7\linewidth]{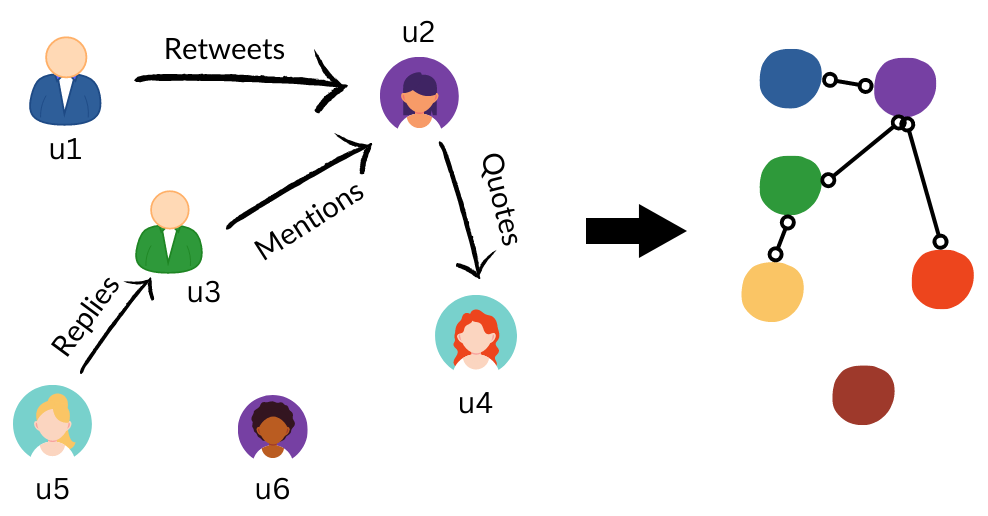}
  \caption{Construction of the user graph}
  \label{fig:graph_creation}
\end{figure}

The data that have been collected can be used to map the interactions between the involved users in the form of a graph. This data consist of pure tweets, retweets, quotes and replies. Pure tweets may include mentions to other users, while all the other types definitely include at least one. 

To form a graph from the collected dataset we proceed as follows: Let us consider a user $u$ that retweets, mentions, replies or quotes another user $u'$. We represent the users $u$ and $u'$ as nodes of a graph; between these nodes/users we assign an edge that represents the bidirectional connection between them. \rev{We (i) consider interactions $u \rightarrow u'$ and $u' \rightarrow u$ equivalent (i.e., undirected graph) and (ii) each type of interactions (retweets, mentions, etc.) is treated equally and contributes to a single undirected edge between users regardless of the number of interactions (unweighted graph).}
\footnote{\rev{Although constructing a directed and weighted graph could provide richer information, the sparsity and variability of our collected dataset would result in poor graph analysis outcomes, as well as we would not be able to apply the polarization and influencer detection algorithms (see Sections~\ref{sec:polarization}. and ~\ref{sec:influencers}).}}

Figure~\ref{fig:graph_creation} depicts an example of a graph formation from interactions in Twitter. To implement the graph generation we used the open-source PyPoll library \cite{pypoll}.

For our analysis, we generate a separate graph for each day. Each graph represents the interactions between users for this specific day\footnote{Similarly, we generate graphs for longer time periods.}. These graphs will be further used to identify the most influential users in \secref{sec:influencers}) and to calculate the polarization between different groups of users in \secref{sec:polarization}. 

\subsection{Political inference}\label{sec:political-inference}
A part of our analysis focuses on the political opinions of the users participating in the discussion and the overall polarization of the discussion. To this end, we attribute to each of the 33K users a political opinion in the following way: (i) For each political party and politician account (see \secref{sec:complementary-datasets}), we collect their followers from the Twitter API. \rev{Here, we remind that each political party and politician account is annotated as "Right", "Left" or "Center" (see details in \secref{sec:complementary-datasets}).} (ii) Then, for each user participating in the discussion of wiretappings, we find how many "Right", "Left" or "Center" accounts they follow. (iii) If a user follows more "Right" accounts, then we attribute their opinion to the political "Right"; and similarly for the "Left". (iv) For users who follow an equal number of Left and Right political accounts, or follow more Central accounts, we annotated them as Central. (v) Finally, users who do not follow any political accounts are defined as neutral.

\rev{\textit{Remark:} To examine the robustness of our political inference methodology, we considered different thresholds for the fraction of "Left"/"Right" follows: e.g., for a threshold 0.75, an account is considered as "Left" if at least 3 out of 4 of the politicians accounts it follows correspond to "Left" politicians, otherwise it is considered "Central" or neutral. While the absolute number of "Left"/"Right" accounts are different for different thresholds, the qualitative results for the polarization in our study do not change. }


\subsection{Polarization detection}\label{sec:polarization}
Polarization is a social phenomenon that has been studied for decades~\cite{isenberg1986group}. The term means that a society is split to two or more groups, based on the opinions of its individuals on a topic, and individuals in each group tend to adopt the views of the group. Many scientists measured that phenomenon in social networks like Twitter~\cite{matakos2017measuring}. 

For our analysis, we quantify the polarization of the discussion about the topic of wiretappings using the Friedkin \& Johnsen (FJ) polarization metric~\cite{matakos2017measuring}; \textit{this metric takes on values in the interval from 0 (no polarization) to 1 (high polarization)}. \rev{The FJ metric measures the degree of opinion polarization within a group. It is based on an opinion dynamics model (Markov Chain), which accounts for both social influence (opinions of neighboring node in the graph) and individuals' resistance to changing their initial opinions. The metric is calculated by assessing the variance in opinions among group members at equilibrium (stationary distribution), reflecting the extent to which opinions are spread out. }

The calculation of the FJ polarization metric is based on the user graph (\secref{sec:graph-generation}) and the political inference of the users (\secref{sec:political-inference}), using the PyPoll library \cite{pypoll}. In the corner case that all the users that are attributed to the political "Right" interact 
only with other "Right" users, and users attributed to the political "Left" interact only with other "Left" users, then the polarization metric would be 1; on the contrary, if every user interacts with the same number of "Right" and "Left" users, then the network would not be polarized. In practice, these corner cases never happen in online discussions, and polarization values are between 0 and 1. 

\subsection{Influencers identification}\label{sec:influencers}
We identify the most influential users for each daily graph based on graph algorithms that rank nodes based on their importance: the higher ranked graph nodes are identified as the most influential users. We tested several algorithms, namely Pagerank, Betweenness Centrality and NetShield~\cite{chen2016node}, and obtained similar results for all of them. 

\rev{We selected to proceed with NetShield due to its efficiency, and relevance to our objectives. Specifically, NetShield operates by optimizing the selection of a subset of nodes (influencers) that, if immunized or removed, would minimize the spread of information or the overall influence in the network. \textit{This aligns with our goal of identifying users who play critical roles in information dissemination}. Finally, it is also able to handle large-scale networks efficiently\footnote{\rev{Computational complexity of NetShield is $O(nk^{2}+m)$, where $n$ is the number of nodes, $k$ is the number of nodes to select, and $m$ is the number of edges.}}, making it ideal for our dataset, which involves extensive interactions and a significant number of users. }

\section{Analysis and Results}\label{sec:analysis}

\begin{figure}
  \includegraphics[width=\linewidth]{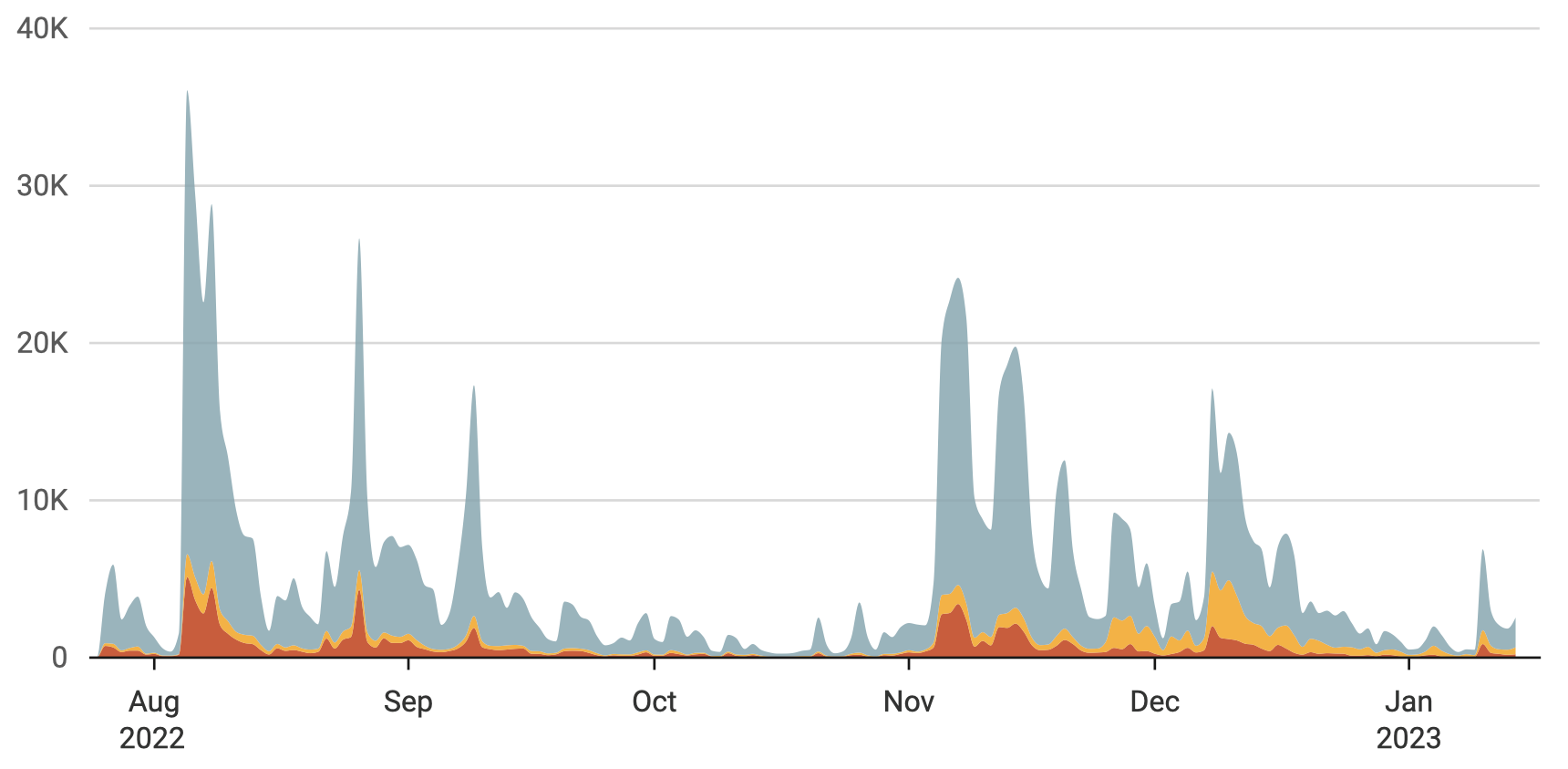}
  \caption{Total number of posts (x-axis) per day (y-axis), grouped by type: tweets (red), retweets (blue), quotes and replies (yellow). The total number of posts between 25-July-2022 and 14-Jan-2022 is 953,722.}
  \label{fig:evolution_posts}
\end{figure}

\subsection{Quantitative analysis}\label{sec:quant-analysis}
We analyze the online discussions for a period of almost a year. Specifically, we calculate the following total and daily statistics:
\begin{itemize}[leftmargin=*,nosep]
    \item number of tweets, hashtags, users, and URLs
    \item most liked, retweeted, and replied tweets
    \item most mentioned, influencing, and active users 
    \item most shared URLs, images, and videos
    \item most popular textual content (words, phrases, and hashtags)
\end{itemize}

All the results of our analysis (daily and aggregate\footnote{We provide date filtering as an option, so that users can explore the aggregate data and statistics that refer to a specific time period.}), as well as interactive visualizations of them, are available through a Web portal~\cite{portal2023ypoklopes}. Here, as an example, we present in Fig.~\ref{fig:evolution_posts} the total number of posts per day, within the time period that the discussion took actively place on Twitter. 

Beyond this quantitative analysis, in the following sections, we dive into specific aspects of the online discussion (for an extended critical interpretation of these results see also~\cite{kiki2023ypoklopes}).

\begin{figure}
  \includegraphics[width=0.7\linewidth]{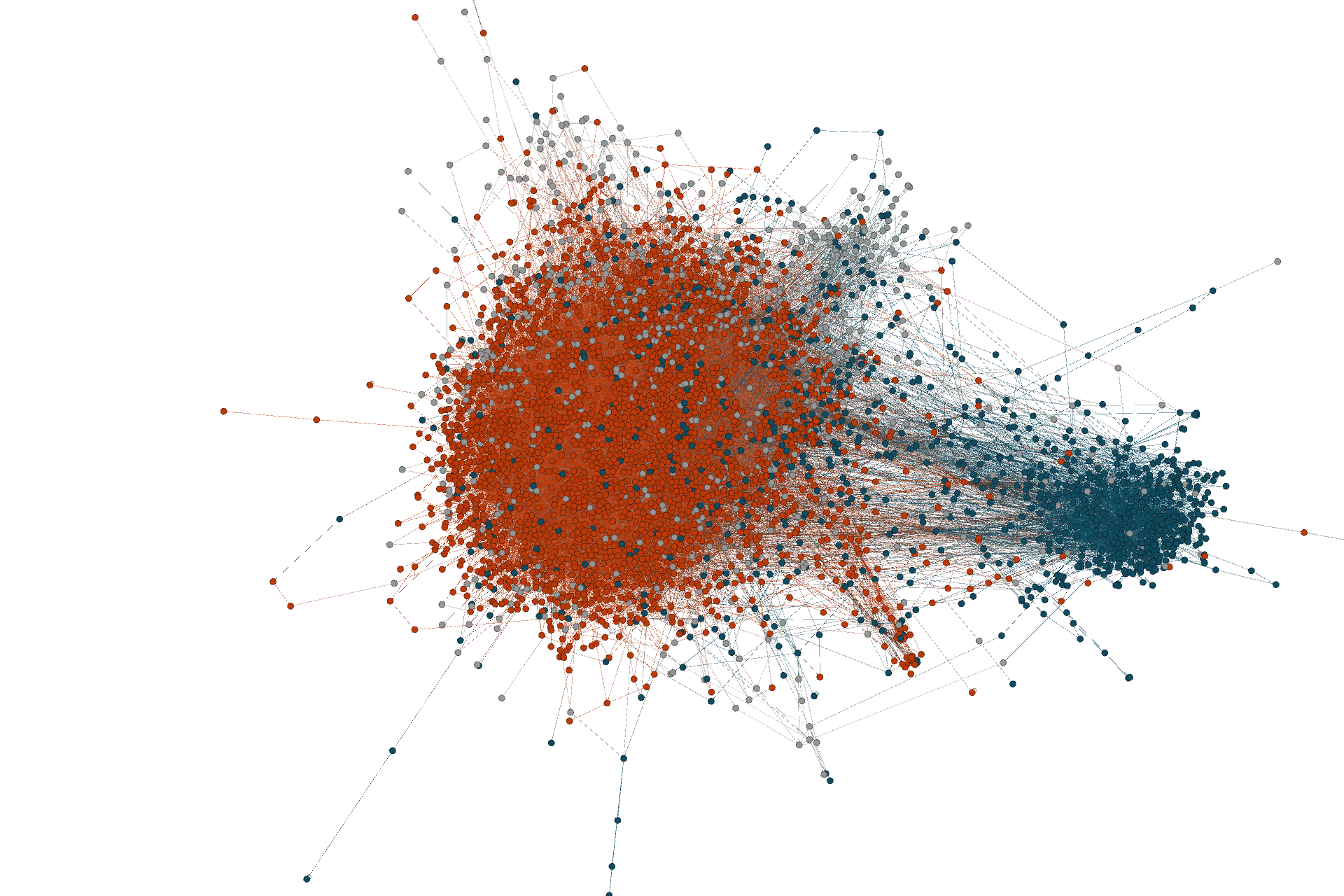}
  \caption{Graph visualization, depicting users attributed to political "Left" (red) and "Right" (blue); August 5, 2022.}
  \label{fig:graph_2022_08_05}
\end{figure}

\subsection{The role of media}\label{sec:media-role}

\myitem{Individuals vs. Media: who drives the discussion?} By analyzing the top 500 accounts, we observe that while 18\% of the total sample are media/journalist accounts, when we focus solely on the 500 users that posted the most tweets, the percentage of media/journalist accounts rises to 35\% and individuals participate by 61\%. On the contrary, over 90\% is the share of individuals, when it comes to the analysis of the 500 accounts that have posted the most replies and quotes. These findings show that a large part of the online discussion is driven by Media, while individual users tend mostly to share opinions rather than creating original content.

This role of Media can be also seen in Fig.~\ref{fig:total_posts} that presents the daily number of tweets over time, and included annotations of the major events and publications related to the wiretappping scandal. We can observe that several of the spikes (i.e., singificant increase in tweets) is correlated with Media publications (see \secref{sec:scandal-history}).

\myitem{Activity of Media on Twitter.} Media play an important role in disseminating the news, in general, and on Twitter, as analyzed above. Exploring the activity of the media/journalist accounts, we find that they mainly post content, rather than interact with other users
. Moreover, the majority 
of the top 20 accounts with the most posted tweets belong to news websites. On the other hand, media rarely get involved in online interactions with other users; e.g., in the list of the top 20 accounts with the higher number of responses to third parties, there is only one journalism group
.

\myitem{Media and Journalists among the top "influencers".} The 11 out of the top 20 accounts identified as the most influencial (see \secref{sec:influencers}) belong to media organizations or journalists. The other accounts to this list, correspond to politicians (e.g., the Greek Prime Minister) or individuals with many followers. Furthermore, looking at the list of the top 20 websites included in tweets, we find that this includes media outlets, the vast majority of which are harshly critical of the government –including those that have been at the forefront of reporting on the wiretapping case.


\subsection{Political opinions: Left vs. Right}\label{sec:left-right}


We conduct a "political profiling" of all users participating in the online discussions for the wiretapping scandal. \textit{Note:} we remind that the government party is classified in the political "Right". Our analysis shows that 70\% of the total \textit{tweets} were posted by accounts that tend to follow mostly left-wing parties and politicians, while only 20\% of the volume of posts seems to come from users that are attributed to the political "Right".


However, the distribution of \textit{users} is not so imbalanced: 42\% are attributed to the political "Left", 29\% to the "Right", 6\% to the "Center", and the remaining 23\% are classified as "Neutral". Combining with the corresponding distribution of tweets analyzed above, this highlights that "Left" users are slightly more (42\%) of users but they “talk” a lot more (70\% of the tweets).


\subsection{Polarization}\label{sec:polarization-analysis}

\myitem{Overall polarization.} We calculate the polarization index (\secref{sec:polarization}) of the discussion on a daily basis, and we find that its values are consistently above 0.5. This indicates a high level of polarization, i.e., users form groups (of similar views) and do not actively interact with other groups. 

An example of this phenomenon is depicted in Fig~\ref{fig:graph_2022_08_05},that shows the user graph (\secref{sec:graph-generation}) for the day of the resignation of the Secretary General of the Prime Minister and the head of NIS (August 5, 2022). This day had the largest amount of total tweets (more than 35K). In the graph, users that are classified as "Left" and "Right" are presented as red and blue dots, respectively. It can be clearly seen that the graph consists of two distinct, weakly connected groups.

\begin{figure}
  \includegraphics[width=0.7\linewidth]{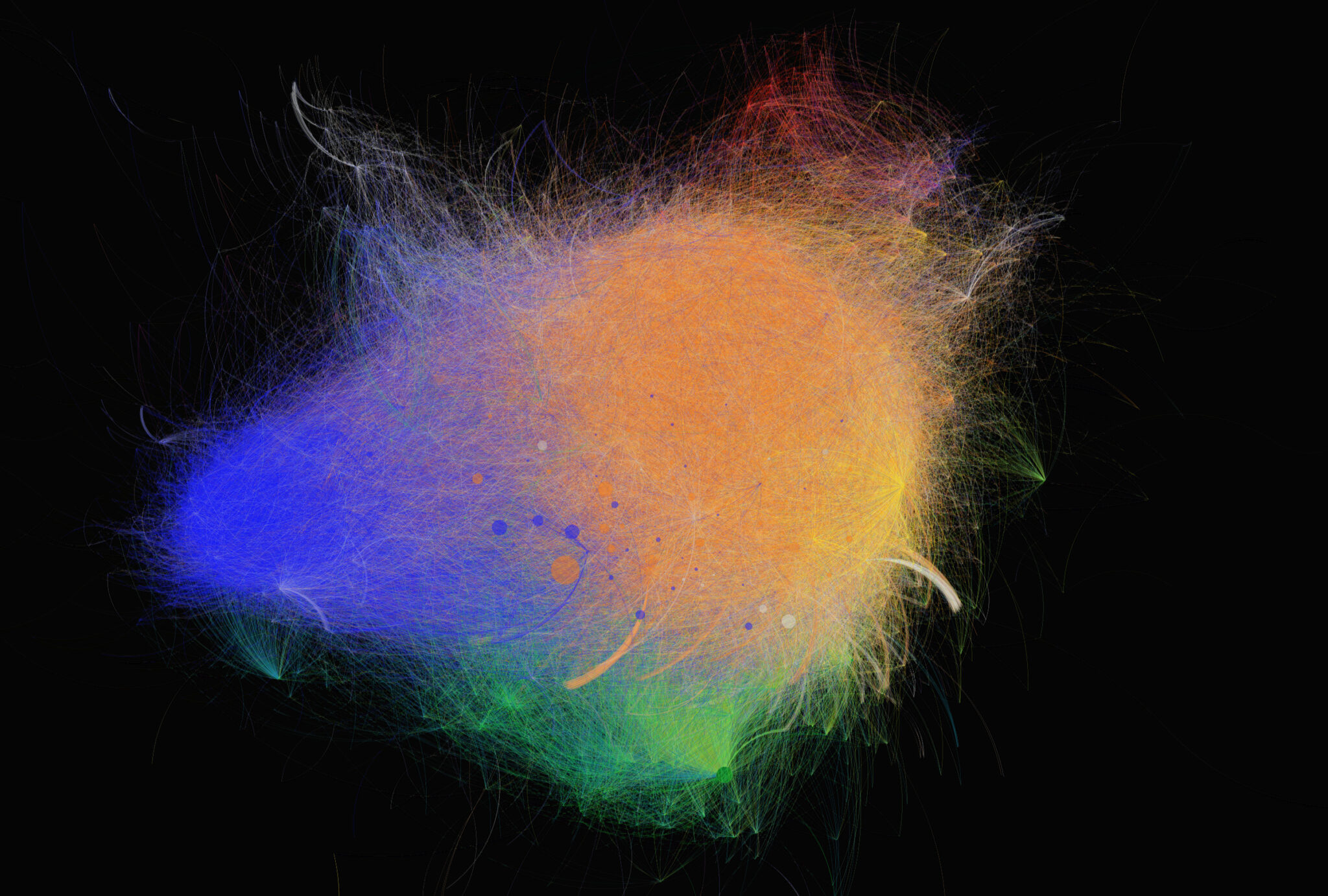}
  \caption{Graph visualization, depicting users attributed to different political parties (colors); Apr. 1, 2022 to Dec. 1, 2022}
  \label{fig:party_graph_pin_light}
\end{figure}


Moreover, Fig.~\ref{fig:party_graph_pin_light} presents the user graph generated based on the data from a 9-month period. Users are colored based on the exact party they are attributed: New Democracy (blue), SYRIZA (orange), MeRA 25 (yellow), KKE (red), and PASOK-KINAL (green). We can see that even when considering these more fine-grained groups, the separation of users based on their political attribution is clear.

\begin{figure}
  \includegraphics[width=1\linewidth]{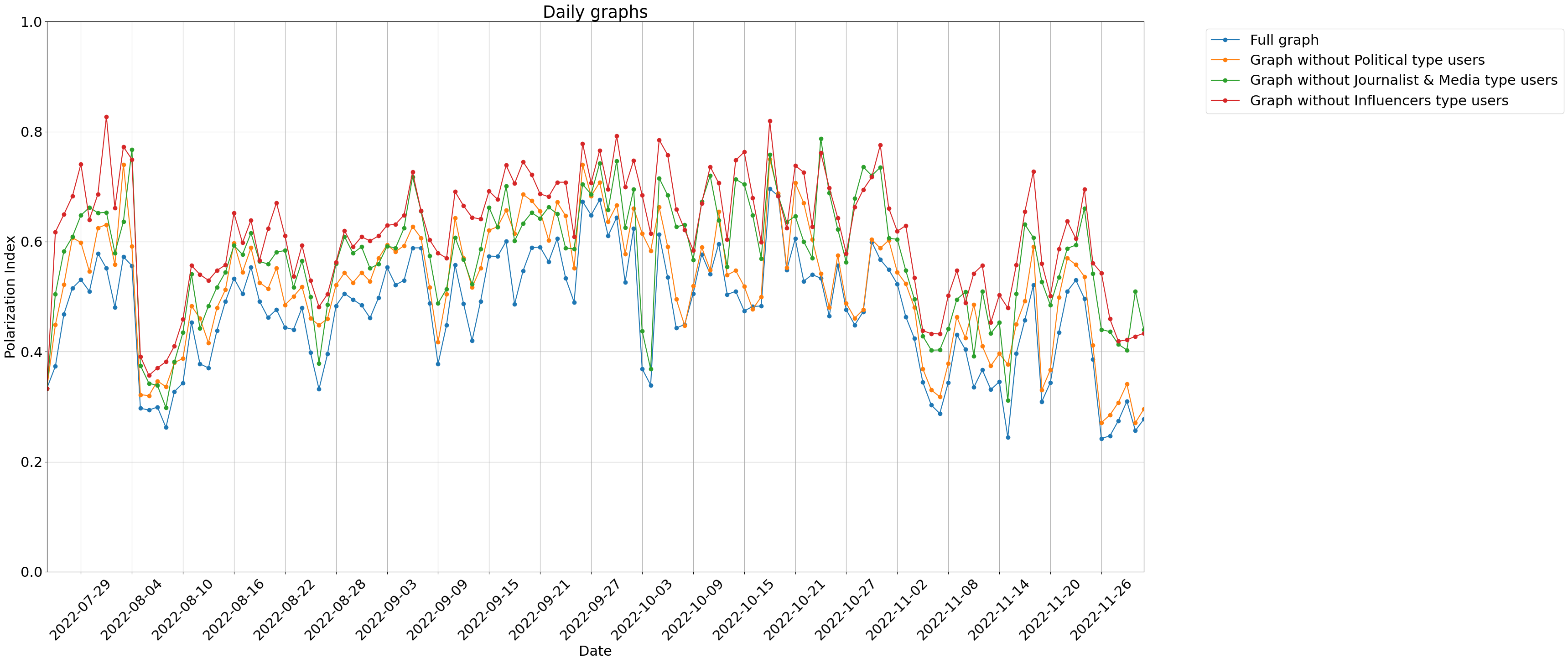}
  \caption{Polarization index (PI) variations over time, and the role of different types of users. }
  \label{fig:daily_graph_polarization}
\end{figure}

\myitem{The effect of different type of users on polarization.} Figure~\ref{fig:daily_graph_polarization} presents with blue line how the value of the polarization index (PI) changes over time. The PI is calculated on the graphs of the tweets/users for each day. We observe that the PI variates around a value of 0.5 with some days having higher PI values (up to 0.7) and some lower values (less than 0.3). These variations depend on the numbers of tweets and users participating in the discussion, but also on who are those users (i.e., their political affiliation), since the discussions on different time periods may focus on different aspects that engage audiences of different political views.

Moreover, in the Fig.~\ref{fig:daily_graph_polarization} presents the daily values of the PI calculated on the graphs without taking into account the users corresponding to Political parties and politicians (orange line), or without Media and journalists (blue line), or without Influencers (red line)\footnote{See Sections~\ref{sec:complementary-datasets} and~\ref{sec:influencers}, respectively, for the details of these users}. We can see that not taking into account the interactions with these users, leads to an increased polarization index; or, in other words, these types of users are the connecting hubs between users of different political views.  Political figures have the least impact, and the largest impact is by omitting Influencers (which is, however, expected due to the definition of these types of users). Noteworthy, the impact of the Journalists and Media is as high as those of all Influences, indicating the role of those types of users in the public discourse around \#ypoklopes

\begin{figure}
  \includegraphics[width=0.5\linewidth]{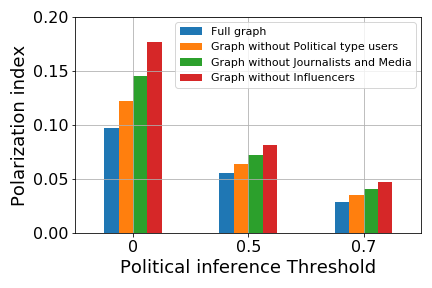}
  \caption{Polarization index vs. threshold for the political inference methodology.}
  \label{fig:pi_vs_th}
\end{figure}

\myitem{Polarization vs. political inference.} Finally, we perform an analysis of the polarization results, by examining the sensitivity of the political inference approach. As discussed in \secref{sec:political-inference}, if a user follows more "Right" accounts than "Left", we assign to it a "Right" political opinion. Here, we apply variations of this political attribution, by assigning an opinion if at least x\% of the accounts followed by a user are "Right" (or, "Left" respectively). E.g., for a threshold $x=50\%$, a user is infered as "Right" if at least half of the accounts they follow are "Right". In Fig.~\ref{fig:pi_vs_th} we present the PI for the entire graph (over all days) for three different thresholds: 0 (i.e., the default) methodology, 50\% and 70\%. We also calculate the same PIs for the graph without the different types of users we analyzed above. As expected, by becoming stricter of how we assign a political inference to a user (i.e., for larger threshold values), the PI reduces. However, the important finding is that the effect of the different types of users remains consistent across all thresholds: Removing political figures, Journalists and Media, or Influencers, always increases the PI, and the effects are similar (qualitatively) no matter what threshold is selected.

\subsection{Communities}\label{sec:communities}
In the concluding phase of our analysis, we implemented a community detection approach on the comprehensive user graph spanning from April to December. Our chosen method is the highly popular Louvain Algorithm \cite{blondel2008fast}, a heuristic approach designed to optimize modularity, a measure of the strength of network community structure. This method iteratively merges nodes to enhance the overall quality of the network partition, efficiently identifying communities by maximizing a quality function rooted in network structure. Notably, the network is constructed based on bilateral interactions among users.

\begin{figure}
  \includegraphics[width=0.49\linewidth]{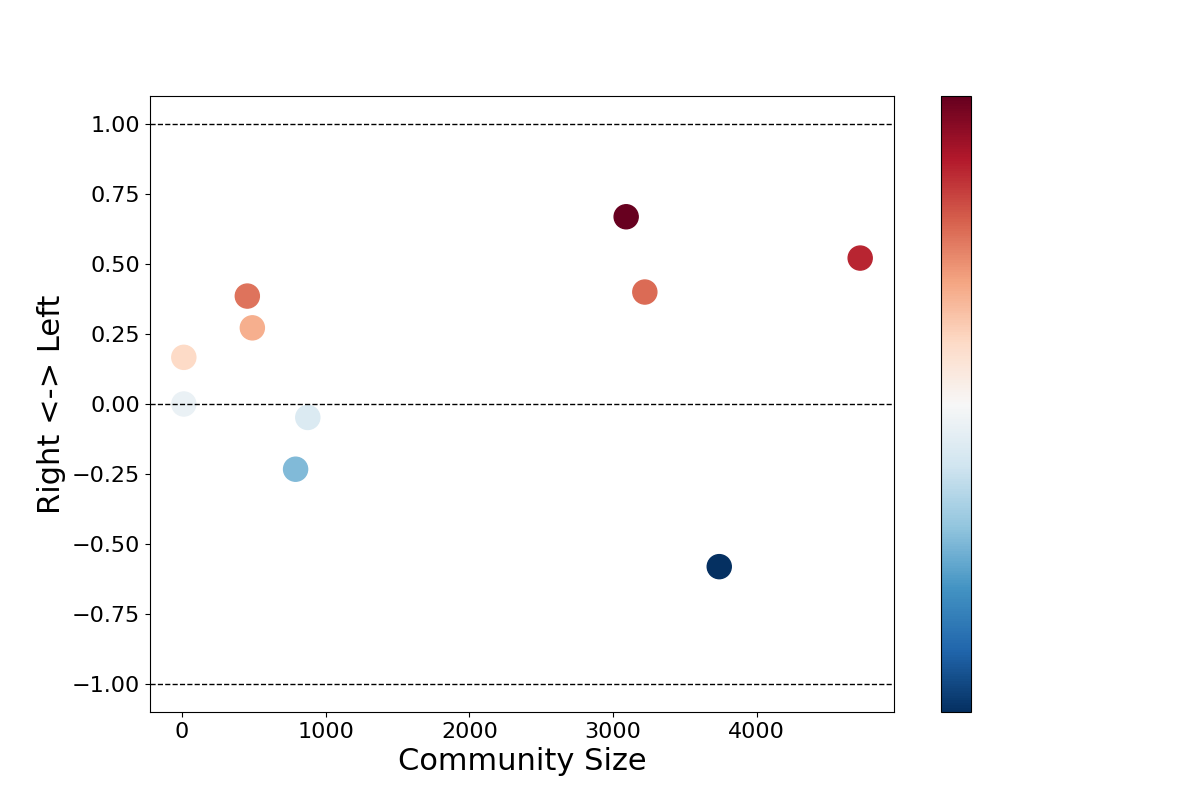}
  \includegraphics[width=0.49\linewidth]{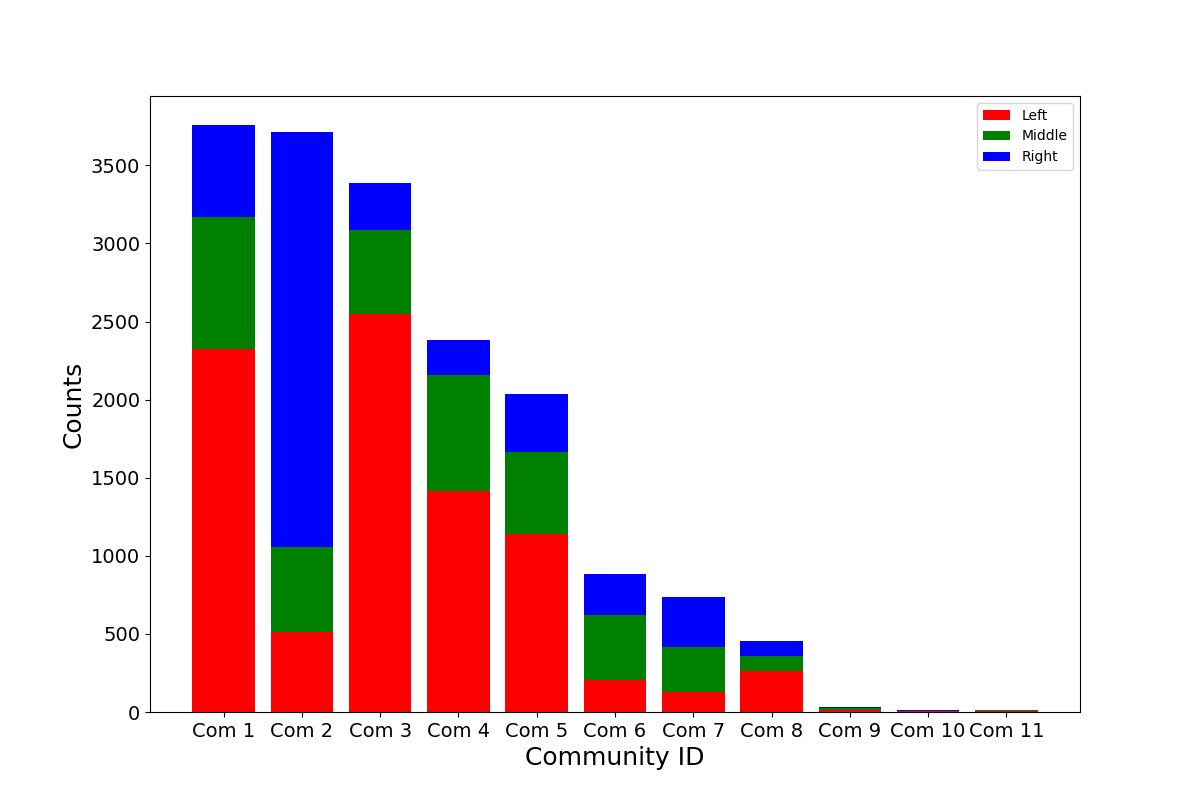}
  \caption{Users communities decomposition based on their political profile.}
  \label{fig:coms}
\end{figure}

The outcome of this analysis uncovers the presence of eight prominent communities, alongside a multitude of smaller ones. We further explore these communities by overlaying them with the "political profiling" methodology previously introduced, which offers insight into the internal composition of each community. The findings, particularly for the top 10 communities based on their size, are visually represented in Figures~\ref{fig:coms}.

Specifically, Fig.~\ref{fig:coms} (left) shows the top-10 communities as circles. The position of the circle in the x-axis indicates the size of the community and the position in the y-axis (as well as the color) indicates the fraction of Right/Left users in the community. We can see that the largest community (around 5000 users) consists of 50\% more Left users than Right. The second largest community is towards Right political opinions. The next two communities (around 3000 users) are both Left-oriented, while the 6 smaller communities are either Left or Right but are less polarized (y-axis values closer to 0).

The same information is presented in Fig.~\ref{fig:coms} (right) that shows the exact percentages of Left (red), Right (blue), and neutral (green) users.


Observations indicate a distinct alignment between each community and a predominant political attribute. This corroborates our earlier observation that users predominantly engage with others sharing similar political beliefs, underscoring the significance of political ideology in shaping online interactions.

\section{\rev{Related Work}}
\rev{Twitter has become an essential platform for political discourse, facilitating large-scale discussions on a wide range of political issues. As presented by~\cite{robertson2019democratic}, who analyzed the democratic role of social media in political debates by examining Twitter activity during the first televised US presidential debate of 2016, Twitter commentary was mainly humorous and negative, but also showed that it played a key role in fact-checking and sharing information, thereby influencing public discourse. In the following, this section reviews recent research on the role of media, bots, polarization and influential accounts in shaping political conversations on Twitter.}

\rev{\textbf{\textit{The Role of Media}}: The influence of media on political discourse on Twitter is significant, as media organizations often act as primary drivers of discussions. \cite{bruns2012researching} highlighted how media organizations use Twitter to engage with audiences and influence political narratives. Their study found that Twitter serves as a platform for news dissemination and public debate, which can shape political narratives. Another study~\cite{alshehhi2019cross} brings out the fact that media outlets played a crucial role in framing discussions and driving public engagement on political and social issues, after analyzing Twitter discussions before, during and after the Ramadan. The impact of news media has also been showcased in a recent research paper~\cite{dagoula2019mapping}, which reveals that political elites and established news organizations maintain their influential positions within Twitter's political conversations, despite the platform’s potential for democratizing public discourse. \cite{dagoula2019mapping} features the different role of media in the challenges of the digital environment, where although they do show high activity, their authority levels have been compromised.} 

\rev{\textbf{\textit{The Role of Bots}}: Bots play a significant role in shaping political discourse on Twitter by amplifying certain narratives and influencing public opinion. \cite{bessi2016social} explored the role of social bots in distorting the online discussion during the 2016 US Presidential election. They discovered that bots were responsible for a substantial number of tweets, thus influencing the overall tone and content of the political discourse on Twitter. Another study~\cite{gallagher2021sustained} examined the amplification of COVID-19-related information by elites on Twitter, noting the significant role bots play in spreading misinformation and polarizing content. Misinformation spreading by bots was also the focus of a study about false information regarding earthquakes in Twitter~\cite{erokhin2023role}. Both these studies underscore the need for effective detection and mitigation strategies to manage bot activities that distort political discourse.} 

\rev{\textbf{\textit{Polarization}}: Polarization in political discourse on Twitter has been a significant area of research, demonstrating how social media can intensify political divisions. \cite{primario2017measuring} investigated the dynamics of polarization on Twitter during the 2016 US presidential election. They emphasized how the platform's features can contribute to polarization by allowing users to participate in political discussions that reinforce their pre-existing beliefs and opinions. These findings were enriched by the results of~\cite{urman2020context}, where authors explored political polarization on Twitter across 16 countries, revealing that the level of polarization varies greatly depending on the local political context and electoral systems. Two-party systems with plurality electoral rules tend to exhibit higher polarization, while multi-party systems with proportional voting show lower levels. The role of media in polarization has also been emphasized by another paper~\cite{gruzd2014investigating}, where the authors used social network analysis to investigate political polarization in Canadian Twitter discourse. They found that media organizations often acted as hubs in the network, connecting disparate user groups and influencing the overall tone and direction of political discussions.}

\rev{\textbf{\textit{Identification of Influential Nodes}}: Understanding the role of influential nodes in Twitter discourse is crucial for comprehending information spread and the impact of key accounts in political discussions. A study~\cite{murthy2016automation} conducted a sociotechnical investigation into how influential users, including media organizations and bots, leverage Twitter to enhance network connections and disseminate information. This research provided insights into how these accounts create social capital and shape political communication through strategic platform use. Another analysis~\cite{tien2020online} focused on the far-right rally in the United States, identifying key influencers and how they propagate information. As presented in this paper, identifying these influential nodes is critical for understanding and potentially mitigating the spread of extremist content.}

\rev{These studies collectively highlight the complex role of media, polarization, bots, and influential accounts in shaping political discourse on Twitter. Our study follows a methodological framework which aspires to address all these challenges effectively. The fact that our results are inline with the main conclusions of previous researchers, further support the importance of media in public political dialogue, especially in Twitter. }

\section{Discussion}

In this work, we studied how a real-world scandal was reflected in online discussions on Twitter. We collected a large corpus of data, analyzed them, and resulted in findings that enhance our understanding of the public opinion about the scandal, and reveal aspects (e.g., polarization) that are difficult to identify with other means. We believe that the methodology we propose can help the investigation of other use cases, and provide useful guidelines and resources for other researchers and journalists.

The key steps for replicating the methodology for a given discussion are the following: (i) The first step is to identify a set of hashtags and keywords that are related to the discussion in hand; this can be done with a manual (journalistic) investigation. (ii) Using available tools and libraries (e.g.,\cite{pypoll}), collect the tweets for a selected period of time. (iii) For identification of Twitter accounts corresponding to political parties and politicians there are available datasets, while for media and journalist accounts a manual annotation is needed (focusing only on the most active accounts, which typically include these kind of users). (iv) For graph generation, political inference, polarization inference and community detection one can use our methodology (\secref{sec:methodology}) that is based on open tools and libraries.

\revv{The findings resulting from this analysis can be used to enrich research and standard journalistic analyses (see a detailed analysis in~\cite{kiki2023ypoklopes}), as well as reveal important insights on the role of social media on public discourse. Specifically:
}
%
\revv{
\\\indent - Similar to the work of \cite{bruns2012researching} and \cite{dagoula2019mapping}, our analysis confirms that media organizations continue to hold considerable influence over the narratives within Twitter discussions. However, our study and methodology also add to this body of research by demonstrating how this influence evolves dynamically in response to real-world events, as evidenced by the intensified discussion patterns we observed in Figure~\ref{fig:total_posts}.
}
\revv{
\indent - In terms of political polarization, our results echo the conclusions of \cite{primario2017measuring} and \cite{urman2020context}, which underscore how Twitter’s platform features can exacerbate political divisions. Our analysis contributes to this understanding, both by revealing how these divisions manifest specifically in the context of scandal-related discourse, but mainly by offering a  perspective (and the methodological steps to accomplish it) on the role of media and influential accounts in either bridging or widening these divides.
}
\revv{
\\\indent - Lastly, in line with previous findings \cite{murthy2016automation} and \cite{tien2020online}, we observed that key influencers (media outlets, political figures, etc.), play a critical role in shaping the online discourse. Our approach to identifying these nodes can be directly applied to future studies aiming to map influence in political discussions on social media.
}

\section*{Acknowledgements}
This research is co-financed by Greece and European Union through the Operational Program Competitiveness, Entrepreneurship and Innovation under the call RESEARCH-CREATE-INNOVATE (project T2EDK-04937).

\begin{adjustwidth}{-\extralength}{0cm}

\reftitle{References}

\PublishersNote{}
\end{adjustwidth}
\end{document}